 \documentclass[final,3p,times,twocolumn]{elsarticle}

\usepackage{amssymb}

\usepackage{lineno}
\usepackage{hyperref}
\usepackage{calc}
\usepackage{soul}
\usepackage{xcolor}

\makeatletter
\def\ps@pprintTitle{%
 \let\@oddhead\@empty
 \let\@evenhead\@empty
 \def\@oddfoot{\centerline{\thepage}}%
 \let\@evenfoot\@oddfoot}
\makeatother

\begin{document}

\begin{frontmatter}



\title{Origin of Enhanced Zone Lines in Field Desorption Maps}

\author[inst1]{Jiayuwen Qi}

\affiliation[inst1]{organization={Dept.~of Materials Science and Engineering, The Ohio State University},
            city={Columbus},
            state={OH},
            country={USA}}

\author[inst1]{Christian Oberdorfer}

\author[inst2]{Emmanuelle A. Marquis}

\affiliation[inst2]{organization={Dept.~of Materials Science and Engineering, University of Michigan},
            city={Ann Arbor},
            state={MI},
            country={USA}}
            
\author[inst1]{{Wolfgang Windl}\texorpdfstring{\corref{cor1}}}
\cortext[cor1]{Corresponding author}

\begin{abstract}
Artifacts in the collective desorption map of the detector hits impede a truthful reconstruction, including enhanced ``zone lines''with high atomic impact intensity. Since APT is destructive, simulation is the only approach to explain the origin of these zone lines, but previous work couldn't reproduce them. Here, we use a new simulation technique that adds the full electrostatic forces to the interatomic forces in a molecular-dynamics simulation and eliminates the previous ad-hoc assumptions. 
We find for the canonical example of tungsten that evaporation happens when the electrostatic force overpowers the interatomic force, and the misalignment of the two forces deviates the launch direction of the atoms in certain zones, giving rise to an accumulation of hit events around zone lines.
\end{abstract}


\end{frontmatter}
Atom probe tomography (APT) is a unique technique able to characterize individual atoms and their 3D arrangement. In Atom probe tomography (APT), individual atoms or molecules are evaporated in high field from a sharp needle shaped specimen, and their impact sequence and position on a 2D detector along with mass analysis allows 3D reconstruction through a simple projection law. 
Desorption or evaporation maps, which are the cumulative images of all atom impacts, often exhibit depleted centers of poles, concentric ring patterns, and depleted and enhanced zone lines connecting poles.
Our ability to reconstruct atomic structures relies on the interpretation of collected APT data, where understanding details in field desorption maps is important to quantify artefacts and improve the reconstruction results. 
While reasonably sounding explanations exist for most of the features observed in field desorption maps, for example, depleted zone lines \cite{vurpillot1999theshape}, the enhanced zone lines have yet to be explained conclusively. The probably most widely believed hypothesis about their origin is the so called ``roll-up'' mechanism proposed by Waugh et al.\ in 1976 \cite{waugh1976investigations}, where the evaporating kink-site atom ``rolls'' up over the plane edge and is channelled by its near neighbors in a specific direction towards the plane center. According to this mechanism, the pole center should receive the most hit events and reveal the highest intensity, which obviously contradicts the depletion of the pole center observed in experiments.  

\begin{figure}
    \centering
    \includegraphics[width=0.9\linewidth]{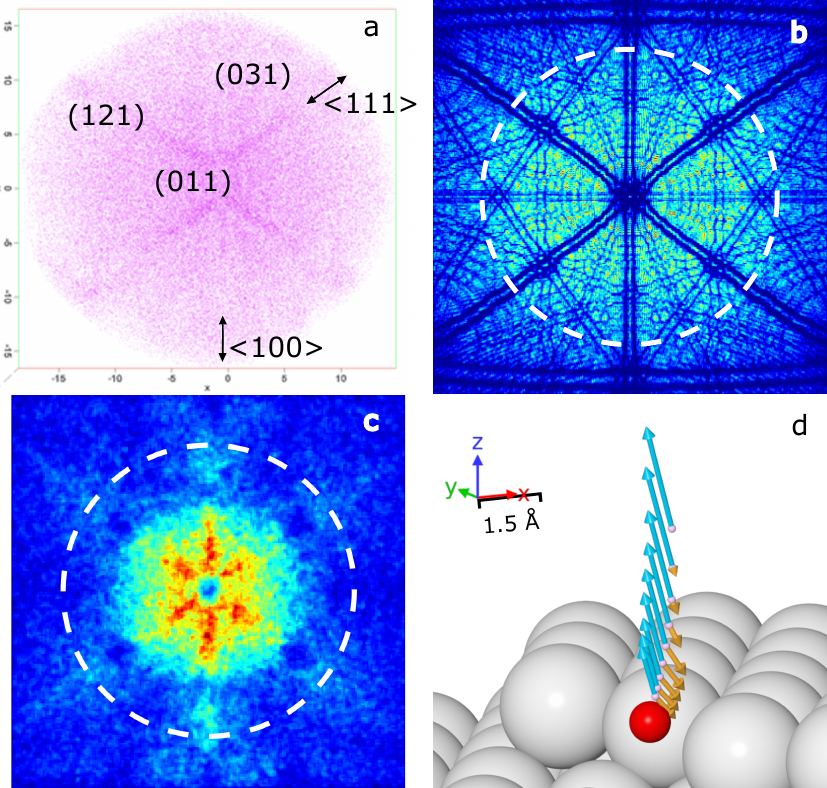}
    \caption{Field desorption maps obtained for $\langle 110\rangle$ oriented W tip in (a) experiment (tip radius = 17 nm); (b) traditional TAPSim simulation (tip radius = 15nm); (c) TAPSim-MD simulation (tip radius = 15 nm). The dashed circles in (b) and (c) indicate a field of view similar to the experimental desorption map in (a). (d) is  an example atom trajectory during field evaporation. The red ball indicates the initial position of the evaporating atom on the tip surface. Small pink balls indicate snapshots of the atom along its trajectory during field evaporation. Blue arrows represent external electrostatic forces, orange arrows represent internal interatomic forces.}
    \label{fig:hit_pattern}
\end{figure}

Since it is impossible to study the formation of desorption maps experimentally due to the intrinsic destructiveness of the field evaporation process, forward modelling has been suggested as the only viable way to investigate this process because each atom is traceable from its site in the virtual tip to the final destination on the detector \cite{vurpillot1999theshape,vurpillot1999trajectories}. However, even the most advanced classical electrostatic methods which have been developed such as the TAPSim method \cite{oberdorfer2013afullscale}, are not able to reproduce enhanced zone lines. In contrast to the experimental desorption map of a $\langle 110\rangle$ tungsten tip in Fig.~\ref{fig:hit_pattern}a, the TAPSim result in Fig.~\ref{fig:hit_pattern}b only shows broad and depleted stripes around zone lines. Similar results have been reported for other classical electrostatic models \cite{vurpillot1999theshape, geiser2009asystem, rolland2015ameshless}. Although state-of-the-art codes like TAPSim include the full physics for determining the electric field and modeling the trajectory to the detector, ad-hoc assumptions are adopted about the moment of evaporation, which are a user-chosen {\it evaporation criterion} determining the atom to be released based on atomic local fields and empirical evaporation fields of species, and {\it zero initial velocity} assigned to evaporated atoms in the trajectory calculation. 

The electrodynamics molecular dynamics simulation approach ``TAPSim-MD'' \cite{method_paper} overcomes this arbitrariness by combining traditional finite element APT simulation  from TAPSim \cite{tapsim,oberdorfer2013afullscale} with classical molecular dynamics as implemented in LAMMPS \cite{lammps,plimption1995fast}. Without making any assumptions about evaporation criteria, atoms of the virtual tip move and get ``evaporated" following the physics of classical molecular dynamics under the net forces composed of electrostatic forces and interatomic forces. Once evaporated (i.e.\ outside the interaction radius with the tip), the further trajectory of the ion to the detector is only determined by the electric field. Based on this approach, TAPSim-MD naturally reproduces enhanced zone lines as we will demonstrate for the canonical example of tungsten. 

Simulation results of three virtual tips with 8 nm, 10 nm and 15 nm in radius are used for our analysis, where the different sizes offer different advantages for specific analyses but fundamentally show analogous results (the test of convergence of the desorption pattern vs.\ tip size has been included in the supplemental materials (see Fig.~S1)). The most detailed desorption pattern is obtained from the largest tip we can simulate at present and thus our displayed desorption map comes from the 15 nm-radius tip. For analyses that require one-by-one inspection and a manageable overall number of atoms, the 8 nm-radius tip is used to collect data for analytical efficiency. 
About 30 layers of atoms (16,270 atoms) are used in the analysis.
For the lateral velocity map, a combination of map clarity and statistical efficiency is necessary, and so we use the 10 nm tip as a compromise for a visually meaningful display, where about 52 layers of atoms are used in the calculation.
 The interatomic potential of tungsten is of the modified embedded atom method type as parameterized by Marinica et al. \cite{wpotential}. 

Figure~\ref{fig:hit_pattern} shows our simulated field desorption map (Fig.~\ref{fig:hit_pattern}c) of the 15 nm W $\langle 110\rangle$ tip in comparison to experiment (Fig.~\ref{fig:hit_pattern}a) and traditional TAPSim simulations (Fig.~\ref{fig:hit_pattern}b). Our TAPSim-MD results clearly reproduce the depleted pole center and enhanced zone lines in agreement with the experiment where the intensity of atom hit events is higher around the $\langle 111\rangle$ and $\langle 100\rangle$ zone lines, while zone lines simulated by the traditional TAPSim approach are depleted with a lower intensity of hit events. The size of the virtual tip with 15 nm in radius is comparable to the experimental tip size of 17 nm in radius, which makes the results directly comparable.

Our correct prediction of the enhanced zone lines makes it now possible to examine their origin. Figure~\ref{fig:hit_pattern}d 
shows an evaporating atom along its initial MD trajectory, along with vectors representing direction and magnitude of external forces (blue) and internal forces (orange). As the atom evaporates away from the surface, the interatomic force decreases while the external force grows. The evaporation is triggered by increasing the applied electric field to a point after which the external force stays stronger than the interatomic force. Other than the traditional approaches, we do not halt the MD at the moment of evaporation and send the atom with zero initial velocity into the trajectory simulator. Instead, the atom keeps its velocity resulting from the forces acting on it. We will show below in detail that for atoms such as the one shown in Fig.~\ref{fig:hit_pattern}d, this velocity includes especially a lateral component that comes from the misalignment of external and internal forces, which deviates atoms towards certain directions and consequently gives rise to enhanced zone lines on the detector map.

\begin{figure*}
    \includegraphics[width=0.9\linewidth]{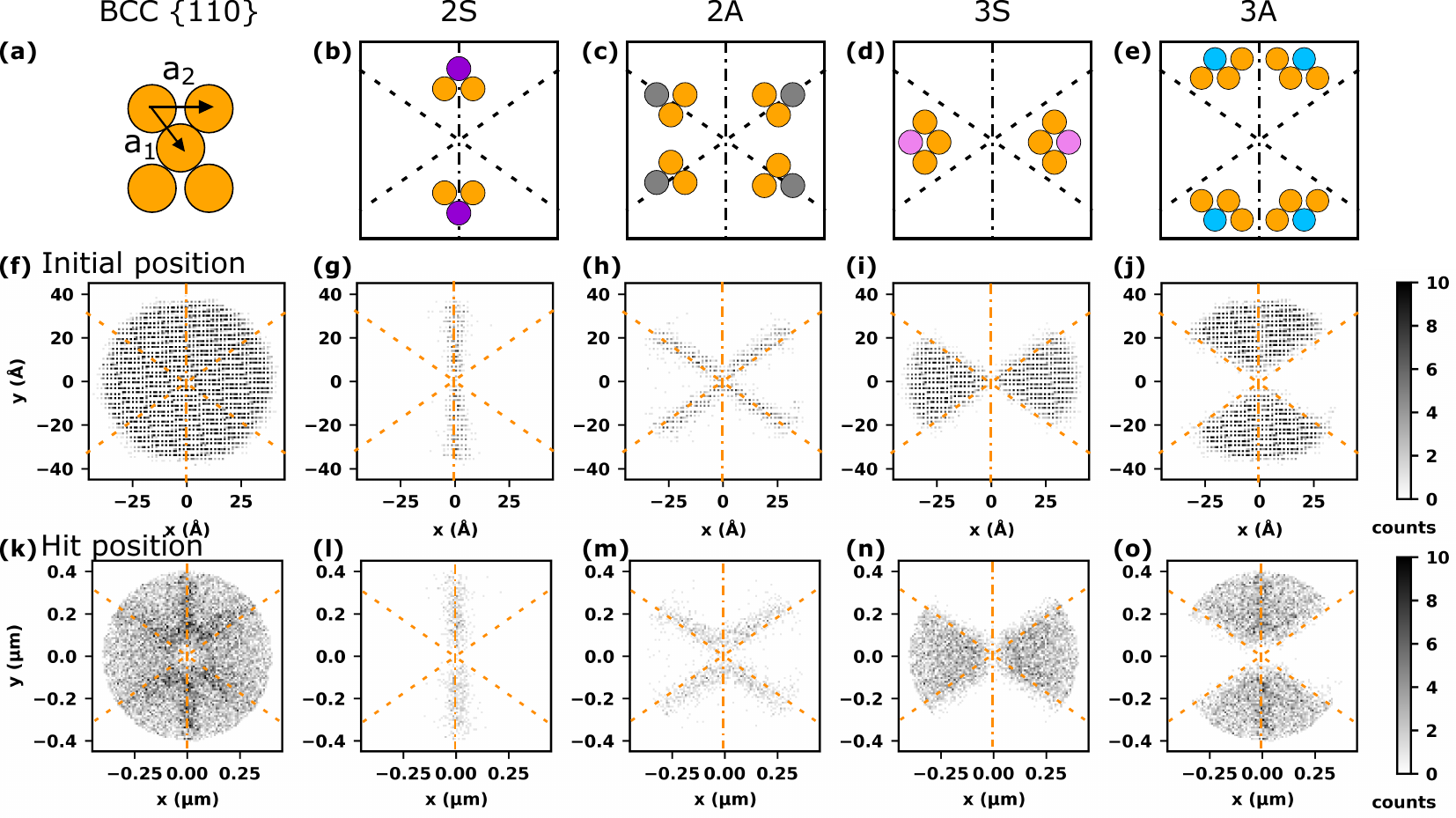}
    \centering
    \caption{(a) is the geometry of BCC {110} plane, where $a_2$ is the lattice constant, $a_1=0.866a_2$; (b)-(e) are 4 classes of atoms categorized by NN configurations (evaporating atoms of each class are colored by purple, gray, pink and blue respectively, neighbor atoms are colored by orange); (f)-(j) are 2D histograms of (x,y) coordinates of all and each class of atoms in the virtual tip (initial positions); (k)-(o) are 2D histograms of detector hit positions of all and each class of evaporated atoms. The intensity of events of 2D histograms are colored on a gray scale. The darker region has a higher intensity of events. Dashed lines represent $\langle$111$\rangle$ zone lines. Dash-dotted lines represent the $\langle$100$\rangle$ zone line. Data are from the simulation with a 8 nm-radius virtual tip.}
    \label{fig:4types}
\end{figure*}

While the view of evaporation as the result of the competition of interatomic has been suggested before \cite{oberdorfer_dissertation}, a common implicit assumption in much of the literature is that the external force is much larger than the internal force. Therefore, the atom should be evaporated very quickly by the huge external force, and the internal force would have little time to make any impact on the atom trajectory. Along with that, it is assumed that the initial velocity component not parallel to the field has negligible effects on the detector image except for some minor blurring \cite{vurpillot2018simulation}.
With regard to the lateral precision and accuracy, the deviation of lateral velocity from the ideal projecting direction is only attributed to trajectory aberrations raised by local field (what we see in TAPSim) or atom motions: diffusion or migration \cite{gault2010influence}, or ``roll-up motion'' \cite{vurpillot2018simulation}.

Based on our findings and common experience in APT experimentation, there are two major arguments why this view should be skewed. First, if the external force was indeed very large compared to the interatomic forces, it should be close to impossible to evaporate a single atom rather than an avalanche of atoms. Indeed, our simulations confirm that the applied voltage has to be merely above the critical value by a small fraction of a Volt to enable atom-by-atom evaporation. This immediately shows that the magnitude of the external and internal force is comparable until the atom leaves the tip surface. Second, in experiments, the voltage is kept significantly below the evaporation voltage and a large number of ``shots'' either with voltage or laser pulses is applied (typically hundreds) before triggering a single atom evaporation event. 

The external force induced by the local electric field is normal to the envelope of the surface and pointing toward the protruding direction. In contrast, the magnitude and direction of the internal force on an atom in MD are determined by the interactions with its nearest neighbors and therefore depend on their configuration. Instead of directly diving into the analysis of forces, we start from an investigation of nearest neighbors of evaporating atoms. The first two nearest neighbor shells (NNs) are taken into account altogether, considering the distance between an atom and its first nearest neighbor $a_1$ is very close to the distance to its second nearest neighbor $a_2$ in a body-centered-cubic (BCC) lattice as shown in Fig.~\ref{fig:4types}a. In addition, we only focus on NNs in the same layer as the evaporating atom, since the layer beneath is always much larger than the surface layer which makes the NN configurations in the layer beneath identical for all evaporating atoms.

As shown in Fig.~\ref{fig:4types}b-e, we categorize evaporated atoms into 4 classes based on the configuration of their NNs just before their evaporation, which are 2S, 2A, 3S and 3A. The number is the count of NNs in the same layer (2 NNs are commonly known as step adatoms, 3 NNs as kink atoms; atoms with NN $>3$ were not found to evaporate in our simulations), and the letter describes the symmetry of the configuration (``S" for symmetric and ``A" for asymmetric). Atoms in the same class have the same NN configuration, which means their internal forces are in equivalent directions. Figure~\ref{fig:4types}g-j show the distribution of initial $(x,y)$ positions of evaporated atoms in each of the classes. It is evident that there is a strong correlation between NN configurations of evaporated atoms and their initial atomic positions in the tip: atoms in 2S are located on the $\langle 100\rangle$ zone line; atoms in 2A are located on $\langle 111\rangle$ zone lines; atoms in 3S are mostly located between two $\langle 111\rangle$ zone lines; atoms in 3A are mostly located between $\langle 111\rangle$ and $\langle 100\rangle$ zone lines. 

The site-dependent NN configuration of evaporated atoms implies a certain sequence of the evaporation events. Atoms with only 2 NNs left before evaporation suggests that they are the last ones evaporated in atomic rows at the edges of the atomic layer, otherwise atoms should have 3 NNs before evaporation. The atoms with 2 NNs are mainly located around $\langle 111\rangle$ and $\langle 100\rangle$ zone lines as indicated in Fig.~\ref{fig:4types}g and h, which implies that atoms located around these zone lines are the last ones to leave the corresponding edges. This simulation result is in agreement with the observation in field-ion images where atoms on $\langle 111\rangle$ and $\langle 100\rangle$ zone lines are commonly observed as evaporation-resistant zone-decoration \cite{waugh1976investigations}. The same conclusion can also be made from the desorption map obtained by the classical TAPSim method without considering any interatomic interaction in Fig.~\ref{fig:hit_pattern}b, where a narrow distribution of atoms is located in the middle of the zone lines. The agreement between the observations from experiment and both simulation approaches suggests that atoms located around zone lines are the last ones to leave, and the evaporation sequence is thus related to the crystallographic features of the material.

More important, since the shape of the atomic terraces stays the same except for shrinking during evaporation, the local NN topology and the resulting directionality of both external and internal force are preserved in the given angular segments throughout the entire evaporation sequence.

\begin{figure}[t]
    \includegraphics[width=0.9\linewidth]{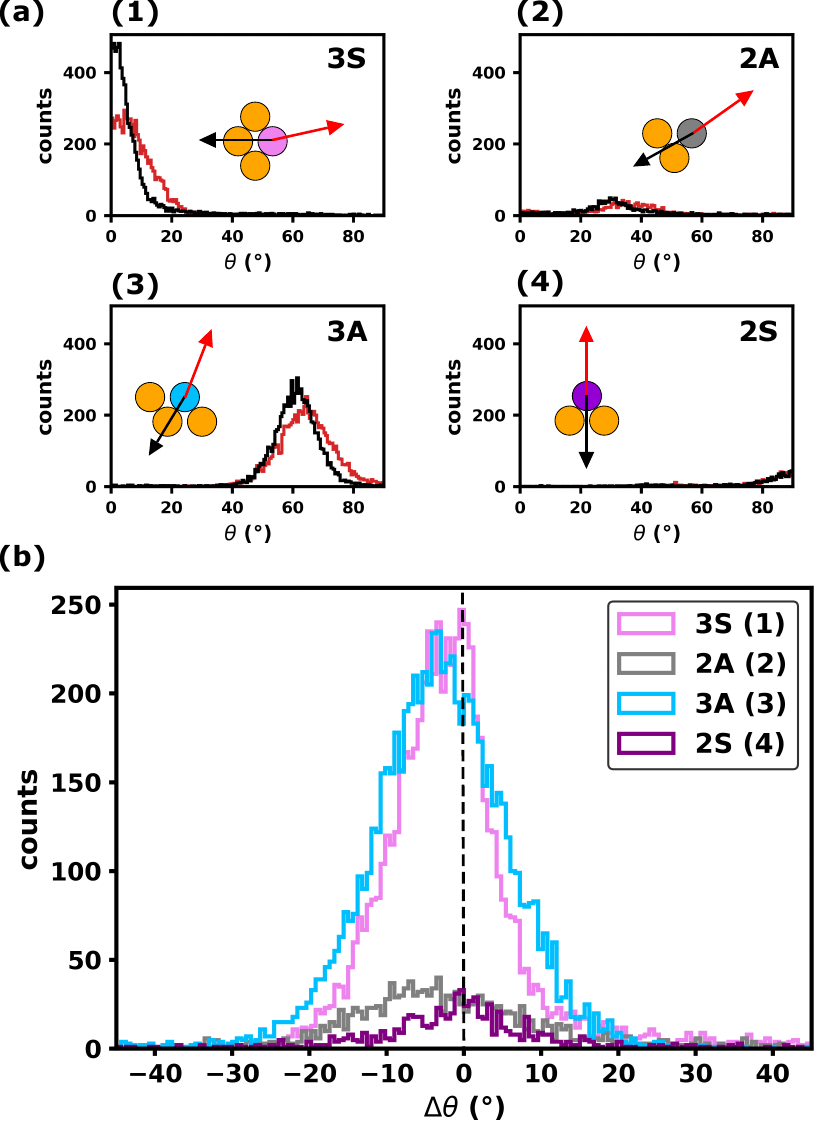}
    \centering
    \caption{(1)-(4) in (a) are histograms of polar angles of the lateral external force and internal force of evaporating atoms in 4 classes. The forces are mapped to the first quadrant of the xy-plane. Polar angle $\theta=arctan(y/x)$ $(0^{\circ}\le\theta\le90^{\circ})$, $(x,y)$ is the force vector in Cartesian coordinates. General force directions are sketched respectively as insets. The evaporating atoms are colored by pink, gray, blue and purple. Neighbor atoms are colored by orange. External and internal forces are represented by red and black arrows respectively. (b) is a histogram of the difference between the polar angles of the two forces mapped to the first quadrant in the same way as in (a), where $\Delta\theta=\theta_{{\rm int}}$-$\theta_{{\rm ext}}$. Data are from the simulation with a 8 nm-radius virtual tip.}
    \label{fig:force_angle}
\end{figure}

With evaporating atoms grouped into four classes by their NNs, we now examine the respective resultant forces. Figure \ref{fig:force_angle}a shows the distribution of the lateral component of external and internal forces felt by atoms in four classes during their evaporation. Because all four quadrants are symmetry equivalent, forces are mapped to the first quadrant of the xy-plane to include more data for statistical analysis. As shown in the sub-panels (1)-(4), the evident correlation between positions of peaks and 4 NN classes demonstrates the topologically preserved directionality of forces of the evaporating atoms in each class. Instead of a uniform distribution, preferences of the external force direction with major peaks around 0 and 60$^\circ$ and minor peaks around 30$^\circ$ and 90$^\circ$ can be explained by faceting of the tip. Equally, the four classes of NN configurations result in a non-uniform angular distribution of the internal force direction with similar but not identical peak positions.

\begin{figure}[h]
    \includegraphics[width=0.9\linewidth]{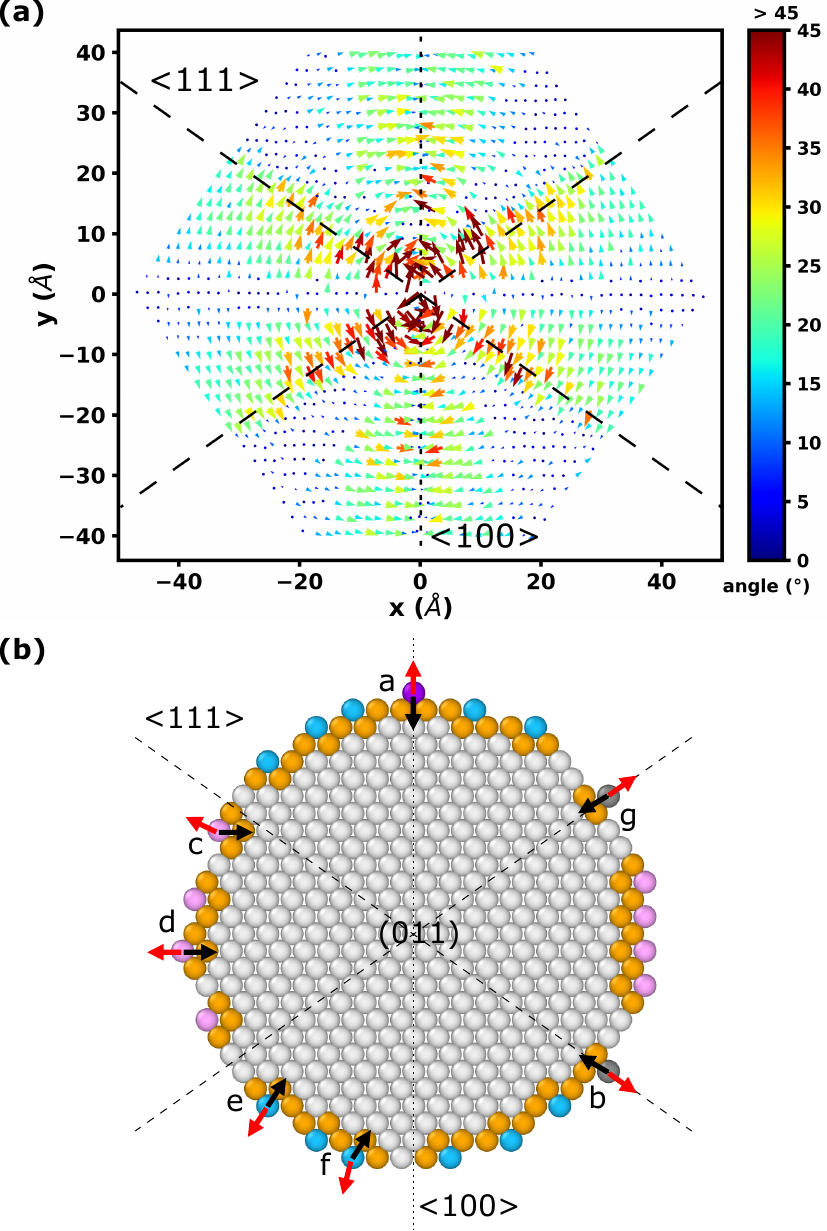}
    \centering
    \caption{(a) is the component of launch velocity transverse to the external force in the xy-plane. Vectors are colored by the angle between the launch velocity and the external force (local electric field). Dashed lines and dotted lines represent $\langle$111$\rangle$ and $\langle$100$\rangle$ zone lines respectively. Each data point is calculated as an average of a column of atoms in a 10 nm-radius virtual W tip. About 52 layers of atoms are used here. (b) is a representative (110) terrace including 4 classes of evaporating atoms. The evaporating atoms are colored by pink, gray, blue and purple. Neighbor atoms are colored by orange. Theoretical directions of lateral external forces and internal forces are sketched for atom a-g, represented by red and black arrows respectively. Dashed lines and dotted lines represent $\langle$111$\rangle$ and $\langle$100$\rangle$ zone lines respectively. }
    \label{fig:dev_v-f}
\end{figure}

Indeed, one can clearly see in Fig.~\ref{fig:force_angle}a that regarding the two main peaks, (1) for 3S and (3) for 3A, peaks of the internal forces shift to the left  of the external force peaks by about 5$^\circ$. To quantify this shift, Fig.~\ref{fig:force_angle}b shows a histogram of the difference between the polar angles of the two forces mapped to the first quadrant as described before, where $\Delta\theta=\theta_{{\rm int}}$-$\theta_{{\rm ext}}$. Zero means the two forces perfectly align, whereas a non-zero shift means a misalignment. Consistent with the left shift of peaks (1) and (3) of the internal forces in Fig.~\ref{fig:force_angle}a, the left-shifted distributions of 3S and 3A in Fig.~\ref{fig:force_angle}b reveal that on average, there is an angle misalignment between the external and internal forces, as the 3S and 3A distributions represent the majority of evaporated atoms. As discussed before, evaporation happens when the external force balances and starts to overcome the internal force, so the resultant total force is close to zero. With that, the transverse force component from the misalignment is not a small distortion of the total net force but large enough to accelerates the atom in a direction deviating from the local field (the ideal assumption).

As shown in Fig.~\ref{fig:dev_v-f}a, this misalignment causes major lateral motion transverse to the local field towards the $\langle 111 \rangle$ and $\langle 100 \rangle$ zone lines, and away from the central (110) pole. For illustration of the link between neighbor arrangement and transverse acceleration, we look at example atoms of the 4 classes of evaporating atoms a representative (110) terrace with typical force alignment in Fig.~\ref{fig:dev_v-f}b. For atom c in class 3S, the misalignment between the two forces gives rise to a component of net force transverse to the external force (local field) pointing towards the $\langle 111 \rangle$ zone line. For atom f in class 3A, the misalignment deviates it towards the $\langle 100 \rangle$ zone line. For atoms a, b, d, e and g, their external and internal forces are well aligned, therefore little deviation is expected. A detailed analysis of the force balance as well as atom trajectory for atom a-g can be found in supplementary Fig.~S2. The comparison between Fig.~\ref{fig:dev_v-f}a and Fig.~\ref{fig:dev_v-f}b shows that the 3S atoms accumulate around the $\langle 111 \rangle$ zone lines on the detector, whereas the 3A atoms enhance the $\langle 100 \rangle$ zone line. This result is also consistent with the 2D histogram of detector hit events in Fig.~\ref{fig:4types}k-o. Indicated by a darker color, the enhanced $\langle 111 \rangle$ zone lines shown in Fig.~\ref{fig:4types}k are a result of overlay of hit events of class 2A (Fig.~\ref{fig:4types}m) and 3S (Fig.~\ref{fig:4types}n), whereas the enhanced $\langle 100 \rangle$ zone lines result from overlay of hit events of class 2S (Fig.~\ref{fig:4types}l) and 3A (Fig.~\ref{fig:4types}o).

In conclusion, we answered the long-standing question about the origin of enhanced zone lines in field desorption maps which cannot be reproduced by previous simulation approaches and leaves explanations to non-quantitatve {\it ad hoc} models such as Waugh's roll-up conjecture \cite{waugh1976investigations}. By using an ``ab-initio'' approach to simulating field evaporation, we successfully reproduce this pattern and can show now for the canonical example of a  $\langle 110\rangle$ oriented tungsten tip that the magnitudes of interatomic forces and field-induced forces are comparable during the evaporation process. The possible misalignment of internal and external forces deviates trajectories of evaporating atoms from local fields and causes accumulations of detector hit events around certain zone lines which explains the observed enhanced zone lines. Similar effects have also been observed in simulations of tungsten in different orientations, which can be found in supplementary Fig.~S3. 

\section*{Supplementary information}
See Supplemental Materials for effects of the virtual tip size, inspection of individual atoms in the simulation, and desorption patterns for tips in different orientations.

\section*{Acknowledgements}

This work is sponsored by AFOSR (PM Dr. Ali Sayir) under Award No. FA9550-14-1-0249 and FA9550-19-1-0378. Computing resources are supplied and maintained by the Ohio Supercomputer Center \citep{OhioSupercomputerCenter1987} under grant number PAA0010. Results of the simulations were visualized with the programs OVITO \citep{ovito} and ParaView \citep{paraview}.

\section*{Competing interests}

The authors declare none.

\pagebreak
\noindent\textbf{\large Supplemental Materials:}

\noindent\textbf{\large Origin of Enhanced Zone Lines in Field Desorption Maps}
\setcounter{equation}{0}
\setcounter{figure}{0}
\setcounter{table}{0}
\setcounter{page}{1}
\makeatletter
\renewcommand{\theequation}{S\arabic{equation}}
\renewcommand{\thefigure}{S\arabic{figure}}
\renewcommand{\bibnumfmt}[1]{[S#1]}
\renewcommand{\citenumfont}[1]{S#1}

\section{Effects of the virtual tip size}

The specimen used in APT experiments usually has a size of several tens of nanometers in radius, however, limited by computational expense, current codes of the enhanced MD method is not able to simulate virtual tips as large as a real specimen. Obviously the simulation is more realistic if a larger tip is used, we need to balance the computational expense and the fidelity. In order to find a proper tip size, we perform a series of simulations in tips with different sizes. The winner is supposed to produce a "converged" desorption pattern with enough details, and in the meanwhile costs a reasonable amount of the computation time.

\begin{figure}[h]
    \includegraphics[width=\linewidth]{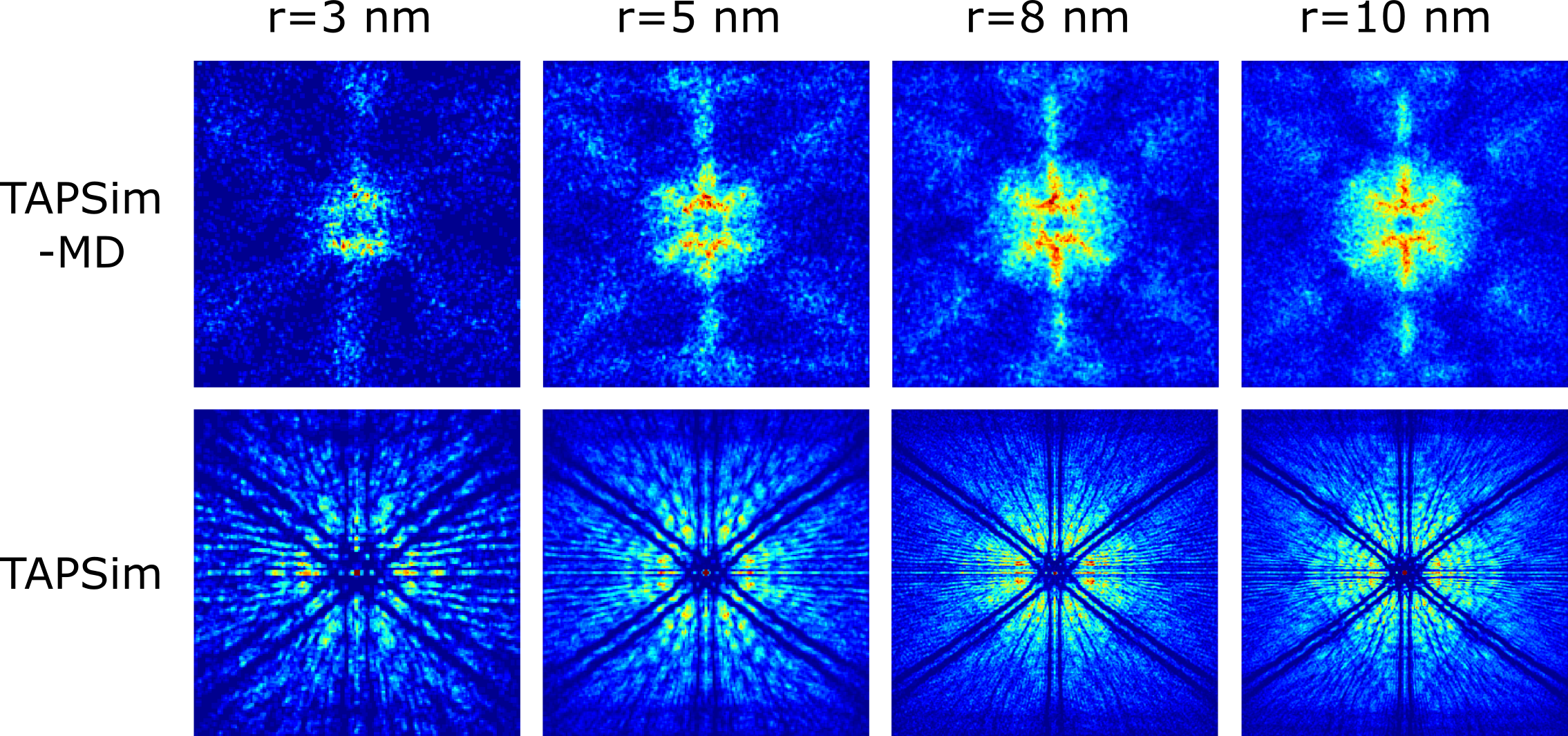}
    \centering
    \caption{Field desorption patterns obtained in virtual tips with different radii. The first row is obtained by the ``MD-TAPS'' method. The second row is obtained by the traditional ``TAPSim'' method.}
    \label{fig:radius_hit}
\end{figure}

As shown in Fig.~\ref{fig:radius_hit}, desorption patterns in the first row are obtained by the "MD-TAPS" method, while the second row is obtained by the traditional electrostatic TAPSim method without MD. Patterns in each column are formed by tips with differen sizes: 3 nm, 5 nm, 8 nm and 10 nm in radius. As observed in simulations, the larger tips reveal more details of the desorption pattern and the pattern tends to converge at 8 nm. Sine the 8 nm tip is able to capture main features of the desorption pattern, for the purpose of efficient data collection and processing, we use data obtained in the tip with this size for most of the analysis in the following sections.

\section{Inspection of individual cases in the schematic plot}

\begin{figure*}
    \includegraphics[width=0.95\linewidth]{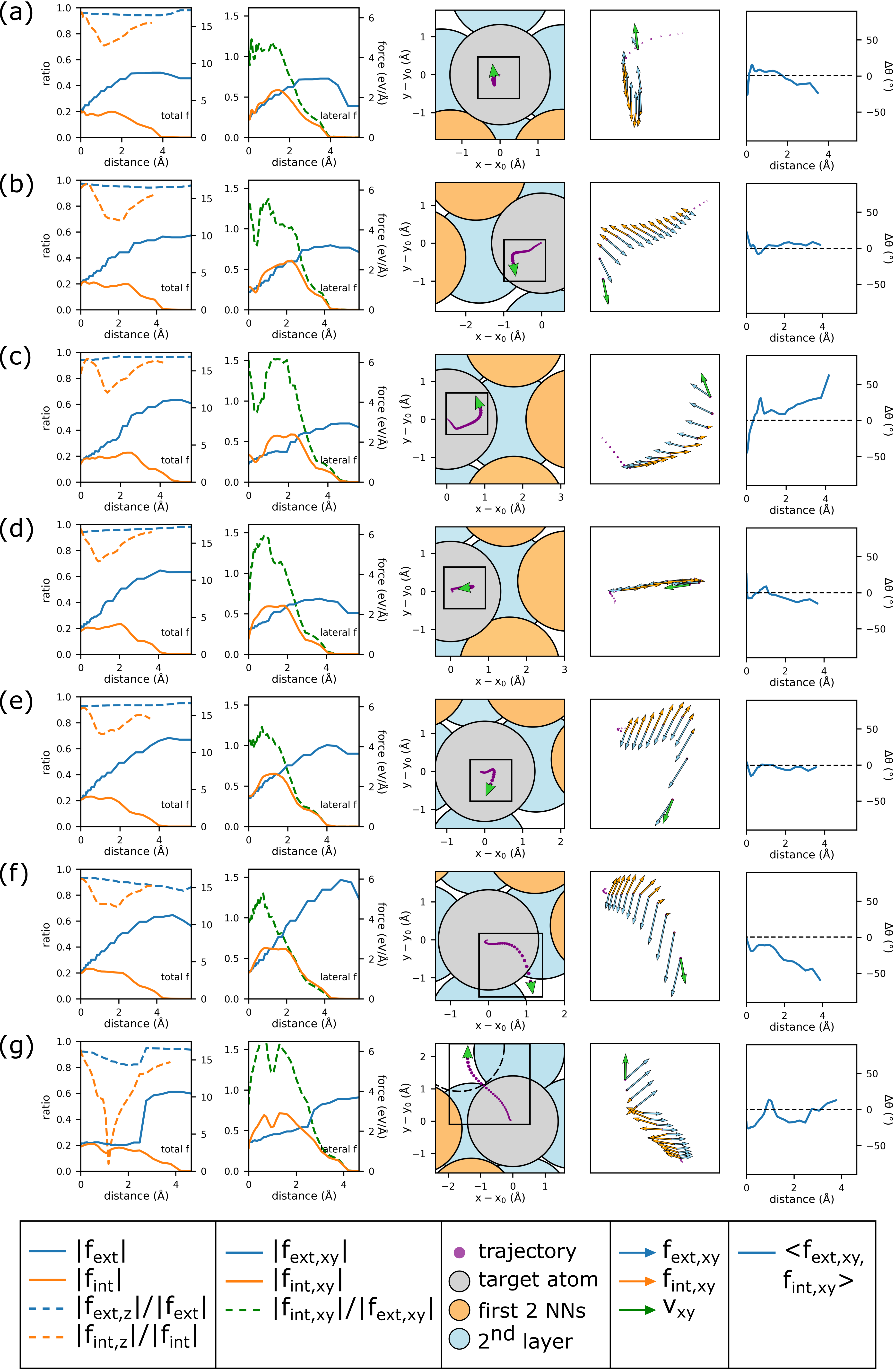}
    \centering
    \caption{Information about force, velocity and trajectory of atom (a)-(g), which are in the similar environment as atom (a)-(g) in the schematic plot fig.~\ref{fig:force_angle}c}
    \label{fig:individual}
\end{figure*}

The representative atoms in the schematic plot Fig.~\ref{fig:force_angle}c have been inspected in the real simulation. As shown in Fig.~\ref{fig:individual}, forces, trajectories and velocities have been investigated in each case. The external force related properties are colored in blue, while the internal force related properties are colored in orange. Before being evaporated atoms always vibrate around their balanced positions in the lattice. Here only moments around evaporation of atoms are focused. The legend for each column of the figure is appended at the end of the figure. The "distance" label for x axis is the distance of an evaporated atom from its balanced position. There are five columns of figures. (1) The first column is plotted to demonstrate that z-component is dominant in both external and internal force during field evaporation. The left axis is the ratio between the magnitude of z-component and the force. The right axis is the magnitude of the force. (2) The second column is to compare the magnitude of the x-y component between the external and internal force. The left axis is the ratio between the magnitude of the x-y component of the external force and that of the internal force. The right axis is the magnitude of the x-y component of each force. The ratio around or higher than 1 demonstrates that the internal force is comparable to the external force in the lateral plane. (3) The third column is a top view of the target atom and its surroundings. The target atom is colored in gray and located at its balanced position. The purple dots show the atom trajectory during evaporation. The first two nearest neighbors in the same layer are colored by orange, and atoms in the layer below are colored by blue. The green arrow shows the direction of the launch direction in the x-y plane. (4) The forth column is a zoom-in plot for the square scope in each of the figures in the third column. The purple dots are still the atom trajectory. The blue arrows and oranges arrows represent the lateral external and internal forces respectively. (5) The last column traces the change of the angle between the direction of the external force and the opposite direction of the internal force in the x-y plane.

Except for case g, all 6 cases are consistent with our postulation. Atoms in case a and b are with 2NNs. The external and internal force in both cases are aligned almost along the same line, and the launch velocity follows the direction of the external force as expected. Atoms in cases c and d are with symmetric 3NNs configuration, however, the location of the atoms in two cases are different. Atom c is located at a higher y position, while atom d is located around the x axis. The direction of the internal force in both cases are similar due to the same NN configuration. The direction of the external force are different due to the distinct geometry. As a result, the launch velocity of atom c is diverted from the field direction, while atom d is evaporated following the field. Similar to case c and d, atoms in case e and f are with asymmetric 3NNs configuration, but the atom f is located closer to the y axis. The external and internal force are aligned with each other very well in case e, while there is a large misalignment in case f. As a result, the launch velocity of atom f has a larger deviation from the field direction.

The result of case g is also informative. One of the neighbors of the target atom g is evaporated just before atom g. The location of this neighbor is indicated by a dashed circle. In this case, the launch direction deviates a lot from the local field, although the local environment is very similar to that in the case b. The analysis of forces reveals that the evaporation of this neighbor causes a sudden change of forces, thus makes atom g evaporate in an unexpected way. It can be told from this case that the consecutive evaporation may have influence on atom trajectories and cause artifacts in APT data. 

\section{Desorption patterns for tips in different orientations}

\begin{figure}[h]
    \includegraphics[width=\linewidth]{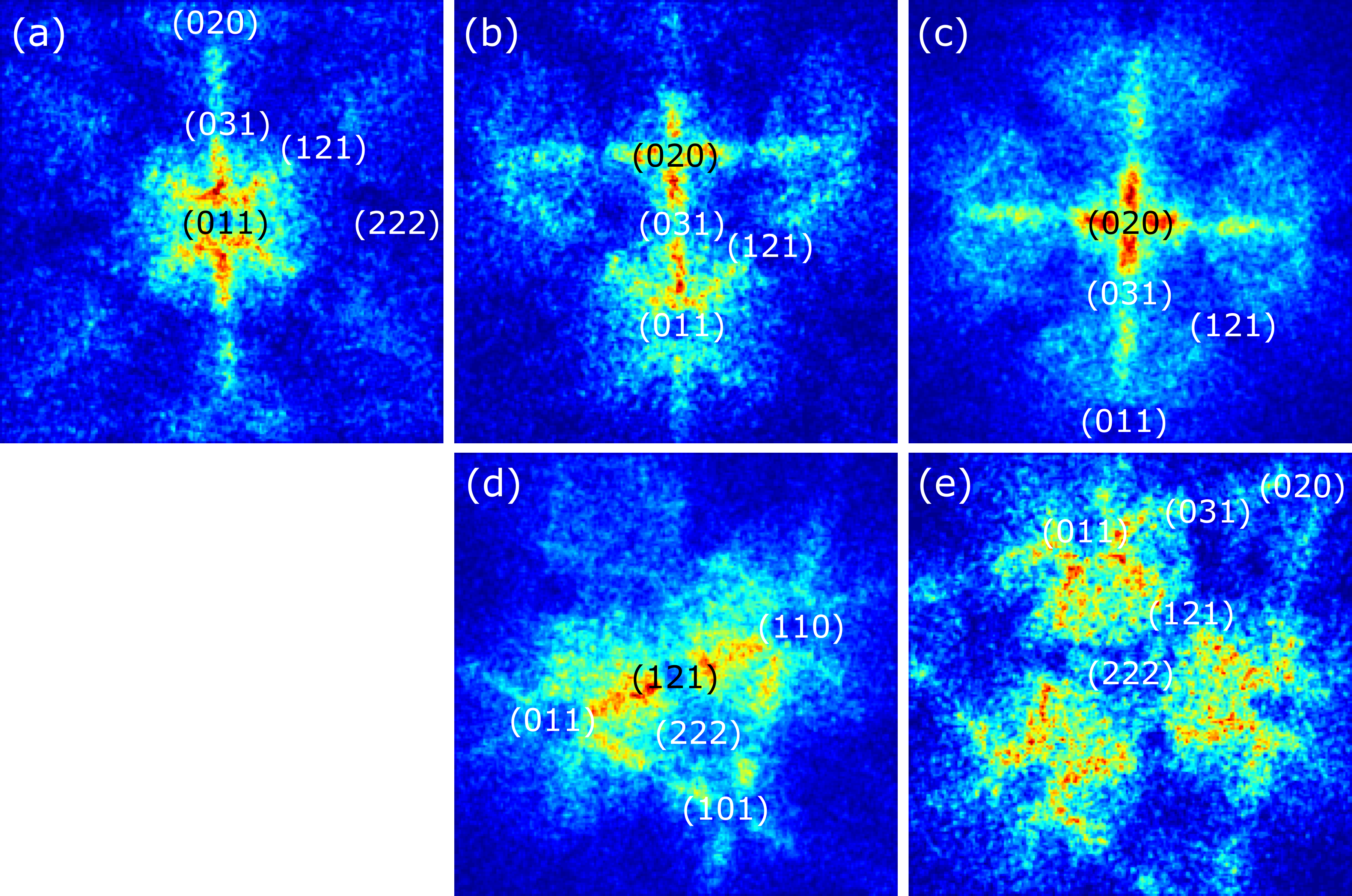}
    \centering
    \caption{Desorption patterns for w tips of size of 8 nm in radius in different crystallographic orientations: $\langle 110\rangle$, $\langle 130\rangle$, $\langle 100\rangle$, $\langle 121\rangle$, $\langle 111\rangle$.}
    \label{fig:hit_orientations}
\end{figure}

Simulations have also been done with 8 nm W tips in different orientations. As shown in Fig.~\ref{fig:hit_orientations}, the desorption patterns obtained for different orientations all have enhanced $\langle 111\rangle$ and $\langle 100\rangle$ zone lines, which is consistent with the experimental results. This consistent feature of enhanced zone lines demonstrates that: (1) the way of facetting and the shape of each facet do not change with the orientation of the tip; (2) atom trajectories are deflected towards zone lines as a result of the misalignment between the electrostatic force and the interatomic force, regardless of the orientation of the tip.

\begin{thebibliography}{10}
\expandafter\ifx\csname url\endcsname\relax
  \def\url#1{\texttt{#1}}\fi
\expandafter\ifx\csname urlprefix\endcsname\relax\def\urlprefix{URL }\fi
\expandafter\ifx\csname href\endcsname\relax
  \def\href#1#2{#2} \def\path#1{#1}\fi

\bibitem{vurpillot1999theshape}
Vurpillot, Bostel, Blavette, The shape of field emitters and the ion
  trajectories in three‐dimensional atom probes, Journal of Microscopy 196
  (1999).

\bibitem{waugh1976investigations}
A.~R. Waugh, E.~D. Boyes, M.~J. Southon, Investigations of field evaporation
  with a field-desorption microscope, Surface Science 61~(1) (1976) 109--142.
\newblock \href {https://doi.org/https://doi.org/10.1016/0039-6028(76)90411-8}
  {\path{doi:https://doi.org/10.1016/0039-6028(76)90411-8}}.

\bibitem{vurpillot1999trajectories}
F.~Vurpillot, A.~Bostel, A.~Menand, D.~Blavette, Trajectories of field emitted
  ions in 3d atom-probe, The European Physical Journal - Applied Physics 6~(2)
  (1999) 217--221.
\newblock \href {https://doi.org/10.1051/epjap:1999173}
  {\path{doi:10.1051/epjap:1999173}}.

\bibitem{oberdorfer2013afullscale}
C.~Oberdorfer, S.~M. Eich, G.~Schmitz, A full-scale simulation approach for
  atom probe tomography, Ultramicroscopy 128 (2013) 55--67.
\newblock \href {https://doi.org/10.1016/j.ultramic.2013.01.005}
  {\path{doi:10.1016/j.ultramic.2013.01.005}}.

\bibitem{geiser2009asystem}
B.~P. Geiser, D.~J. Larson, S.~Gerstl, D.~Reinhard, T.~F. Kelly, T.~J. Prosa,
  D.~Olson, A system for simulation of tip evolution under field evaporation,
  Microscopy and Microanalysis 15~(S2) (2009) 302--303.
\newblock \href {https://doi.org/10.1017/S1431927609098298}
  {\path{doi:10.1017/S1431927609098298}}.

\bibitem{rolland2015ameshless}
N.~Rolland, F.~Vurpillot, S.~Duguay, D.~Blavette, A meshless algorithm to model
  field evaporation in atom probe tomography, Microscopy and Microanalysis
  21~(6) (2015) 1649--1656.
\newblock \href {https://doi.org/10.1017/S1431927615015184}
  {\path{doi:10.1017/S1431927615015184}}.

\bibitem{method_paper}
J.~Qi, C.~Oberdorfer, E.~A. Marquis, W.~Windl,
  \href{https://arxiv.org/abs/2207.03958}{Ab-initio simulation of field
  evaporation} (2022).
\newblock \href {https://doi.org/10.48550/ARXIV.2207.03958}
  {\path{doi:10.48550/ARXIV.2207.03958}}.
\newline\urlprefix\url{https://arxiv.org/abs/2207.03958}

\bibitem{tapsim}
{APT} {S}oftware {TAPSim},
  \url{https://www.imw.uni-stuttgart.de/mp/forschung/atom_probe_RD_center/software/},
  accessed: 2021-09-19.

\bibitem{lammps}
{LAMMPS} {M}olecular {D}ynamics {S}imulator, \url{http://lammps.org}, accessed:
  2021-09-19.
\newblock \href {https://doi.org/10.5281/zenodo.3726416}
  {\path{doi:10.5281/zenodo.3726416}}.

\bibitem{plimption1995fast}
S.~Plimpton, Fast parallel algorithms for short-range molecular dynamics,
  Journal of Computational Physics 117~(1) (1995) 1--19.
\newblock \href {https://doi.org/https://doi.org/10.1006/jcph.1995.1039}
  {\path{doi:https://doi.org/10.1006/jcph.1995.1039}}.

\bibitem{wpotential}
M.~Marinica, L.~Ventelon, M.~R. Gilbert, L.~Proville, S.~L. Dudarev, J.~Marian,
  G.~Bencteux, F.~Willaime,
  \href{https://doi.org/10.1088/0953-8984/25/39/395502}{Interatomic potentials
  for modelling radiation defects and dislocations in tungsten}, Journal of
  Physics: Condensed Matter 25~(39) (2013) 395502.
\newblock \href {https://doi.org/10.1088/0953-8984/25/39/395502}
  {\path{doi:10.1088/0953-8984/25/39/395502}}.
\newline\urlprefix\url{https://doi.org/10.1088/0953-8984/25/39/395502}

\bibitem{oberdorfer_dissertation}
C.~Oberdorfer, Numeric simulation of atom probe tomography, Ph.D. thesis,
  Westf\"{a}lischen Wilhelms-Universit\"{a}t M\"{u}nster (2014).

\bibitem{vurpillot2018simulation}
F.~Vurpillot, S.~Parviainen, F.~Djurabekova, D.~Zanuttini, B.~Gervais,
  Simulation tools for atom probe tomography: A path for diagnosis and
  treatment of image degradation, Materials Characterization 146 (2018)
  336--346.
\newblock \href {https://doi.org/https://doi.org/10.1016/j.matchar.2018.04.024}
  {\path{doi:https://doi.org/10.1016/j.matchar.2018.04.024}}.

\bibitem{gault2010influence}
B.~Gault, M.~Müller, A.~La~Fontaine, M.~P. Moody, A.~Shariq, A.~Cerezo, S.~P.
  Ringer, G.~D.~W. Smith, Influence of surface migration on the spatial
  resolution of pulsed laser atom probe tomography, Journal of Applied Physics
  108~(4) (2010) 044904.
\newblock \href {https://doi.org/10.1063/1.3462399}
  {\path{doi:10.1063/1.3462399}}.

\bibitem{OhioSupercomputerCenter1987}
O.~S. Center, \href{http://osc.edu/ark:/19495/f5s1ph73}{Ohio supercomputer
  center} (1987).
\newline\urlprefix\url{http://osc.edu/ark:/19495/f5s1ph73}

\bibitem{ovito}
A.~Stukowski,
  \href{https://doi.org/10.1088/0965-0393/18/1/015012}{Visualization and
  analysis of atomistic simulation data with {OVITO}{\textendash}the open
  visualization tool}, Modelling and Simulation in Materials Science and
  Engineering 18~(1) (2009) 015012.
\newblock \href {https://doi.org/10.1088/0965-0393/18/1/015012}
  {\path{doi:10.1088/0965-0393/18/1/015012}}.
\newline\urlprefix\url{https://doi.org/10.1088/0965-0393/18/1/015012}

\bibitem{paraview}
U.~Ayachit, The ParaView Guide: A Parallel Visualization Application, Kitware,
  Inc., Clifton Park, NY, USA, 2015.

\end{thebibliography}
\end{document}